# Queering the ethics of AI[1]

Eduard Fosch-Villaronga & Gianclaudio Malgieri

eLaw Center for Law and Digital Technologies, Leiden University, Netherlands

**Abstract**

This book chapter delves into the pressing need to "queer" the ethics of AI to challenge and re-evaluate the normative suppositions and values that underlie AI systems. The chapter emphasizes the ethical concerns surrounding the potential for AI to perpetuate discrimination, including binarism, and amplify existing inequalities due to the lack of representative datasets and the affordances and constraints depending on technology readiness. The chapter argues that a critical examination of the neoliberal conception of equality that often underpins non-discrimination law is necessary and cannot stress more the need to create alternative interdisciplinary approaches that consider the complex and intersecting factors that shape individuals' experiences of discrimination. By exploring such approaches centering on intersectionality and vulnerability-informed design, the chapter contends that designers and developers can create more ethical AI systems that are inclusive, equitable, and responsive to the needs and experiences of all individuals and communities, particularly those who are most vulnerable to discrimination and harm.

# Introduction

---

AI technologies are being rapidly integrated into various sectors of society, from industry to healthcare. These systems can increase productivity and resource efficiency, due in part to the harvesting of vast amounts of data that can be processed extremely fast and predict probable outcomes. However, using past data to generate probable futures may replicate scenarios that society no longer considers desirable. Research shows that AI algorithms display biases towards certain genders, ages, races, and sexual orientations, which can result in harmful outcomes for large parts of society (Xenidis & Senden, 2019; Fosch-Villaronga et al., 2021). For instance, facial recognition systems may struggle to recognize dark-skinned women (Buolamwini and Gebru, 2018; Gebru, 2020), and content moderation tools may incorrectly flag drag queens' language as toxic (Thiago, Marcelo & Gomes, 2021). These biases are often the result of limited datasets that fail to fully represent the society or the systematic configuration biases within the AI scientific community. Like many other fields, the AI community struggles to account for diversity and address inequality. The underrepresentation of women, people of color, and LGBTQ+ individuals in research labs and leadership roles results in a lack of diversity and inclusion considerations in AI development.

While much attention has been paid to issues such as how automation may replace the human workforce and privacy and data protection, there is also a need to consider how AI intersects with issues of gender, sexuality, and identity (Fosch-Villaronga & Poulsen, 2022). Queering the ethics of AI requires us to examine how these technologies can perpetuate or challenge discrimination and oppression and to develop strategies for building more inclusive and equitable systems.

After this introduction, the second section—"Complex and intersecting factors shaping individuals' experiences of discrimination in AI"—expands on some of the inadvertent challenges of AI, its regulation and the ethical configurations that unfortunately support discrimination and bias. And the third and final section "Working towards more inclusive AI practices, methods, and approaches" explores how integrating technical anticipatory methodologies with broader social and ethical considerations is crucial for ensuring that AI technologies serve all individuals' diverse needs and aspirations while minimizing harm and fostering a more just and inclusive society.

# Complex and intersecting factors shaping individuals' experiences of discrimination in AI

**The dualism nature of engineering practice reinforces binarism**

Classifying things as opposites has been deeply ingrained in many societies for centuries. For instance, pairs, doubles, and visual oppositions are prominent in several art forms in most cultures in Andean prehistory, e.g., bichrome silver and gold depicting the moon and the sun, representing female and male figures, are spread out in pre-Inca societies. According to Bernier (2009), these representations reflect the importance of symbolic duality in religions, ritual performances, and social order and have impacted our way of thinking throughout history. This dualistic way of thinking has impregnated many constructions throughout history, including legal systems, across the globe. In engineering, there is also a general inclination for using conceptual dichotomies, like classified/unclassified, yes/no, male/female, concrete/abstract, and reductionist/holistic. This way of organizing things, Faulkner (2000) explains, is ineradicable of the engineering practice given that engineers, in the quest for certainty, usually operate in terms of binary oppositions—i.e., "it works/it does not work," "the light is on/the light is off," etc. For this reason, engineering is based on "a central paradox in which certainty/order/controllable is juxtaposed with uncertainty/chaos/uncontrollable" (Faulkner; 2000, p. 781). The world, however, cannot be (or seems to resist being) simplified into two distinct, often opposite and reductionist, categories (Bucciarelli, 1994; Johnston, 2018).

These "false dichotomies" often have negative consequences when applied to AI. For example, if an AI system is trained to classify people as either male or female, the algorithm may not accurately recognize or include people who identify as non-binary or gender non-conforming (Hamidi et al., 2018). It may exclude the intersex community or misclassify the transgender population (Keyes, 2018). Misgendering can have various consequences depending on the application context and the person involved. In social media, for instance, it may lead, at least, to receiving adverts geared to a different audience. However, this practice can also lead to exclusion, discrimination, and harm (Howanski et al., 2021; Fosch-Villaronga et al., 2021). In more sensitive contexts like medicine, an error or misclassification may lead to a misdiagnosis, which can have fatal consequences for the individual (Cirillo et al., 2020; Fosch-Villaronga et al., 2022). Coupled with the practical mindset of engineering, this problem results from the traditional understanding of concepts such as sex and gender, which are typically reduced to a binary opposite: masculine vs. feminine (Nielsen et al., 2021), especially in the so-called gender classifier systems (Rai, & Khanna, 2012). As one can imagine, these dualistic practices can further entrench discrimination and inequality (Buolamwini & Gebru, 2018), especially because there is, with sex, three types (male, female, intersex) and with gender, many more that may or not correspond to the sex that had been assigned at birth.

More worrisome is the fact that the engineering and medical communities (The EUGenMed et al., 2015) often confuse these two terms even though they have different meanings. Gender is a person's internal sense of their identity; sex is the assigned gender at birth based on medical factors (e.g., genitalia, chromosomes, and hormones). The most crucial difference between these concepts is that sex is assumed to be determined objectively, whereas gender is inherently subjective to the person. Consequently, inferring gender from apparently objective means may be prone to errors (Fosch-Villaronga et al., 2022). Simply put, a system may be wrong if it assumes information that has not been disclosed directly. This means that any form of automated inferences about gender is potentially problematic or harmful for the individuals involved.

But the problem is not only when AI systems automatically attribute gender to an individual; it is also operative when well-meaning human agents intervene in response to these mistakes. Consider, for example, how transgender people experience security control at airports (Costanza-Chock, 2020). Body scanners are usually based on a binarist design paradigm—man/woman—thus systemically excluding transgender individuals by design. Upon a flagged error or misalignment with such a binary system, the intervention of human police officers often follows. However, these human agents often lack appropriate training in diversity and are, therefore, unprepared to deal with the complexities of gender identity, which can exacerbate feelings of distress and discomfort by the affected individual. Fortunately, there is increasing attention now being paid to diversity and inclusion in police training. However, recent research shows that although the training provides expanded knowledge on these topics, they fail to impact strategic approaches, which may indicate that current diversity training methods are unlikely to bring about significant changes in the behavior of law enforcement personnel at all (Lai & Lisnek, 2023).

These examples demonstrate how "human intervention," usually considered an essential ethical safeguard for using and deploying AI, could be ineffective or even harmful. Coupled with many other examples facing intersectional considerations, these examples indicate the need for a paradigm shift in AI ethics—what we call a "queering" of AI ethics—rather than the typical search for simple and quick safeguards. Queering the ethics of AI means reconceptualizing AI ethics with a more open approach, one which can overcome top-down categorization and be more accessible to diverse subjectivites and layered perspectives.

**Technological affordances and constraints affect communities differently**

Both the embodiment of a particular technology (be it a robot for instance that can pick a glass from a table or a user interface in a software application) and its abilities (to what extent a user can perform a pre-conceived task or a new one) afford and constraint user behavior (Majchrzak & Markus, 2013).

A specific area of concern for the Ethics of AI is emerging socially complex virtual environments based on data-intense AI systems. An important example is the metaverse, meant in the broad sense as any possible form of augmented or virtual reality, generally facilitated by wearable technologies (Burrows, 2022). The metaverse has the potential to afford and empower users to self-identify and be seen in ways that could break down social, economic, and physical barriers (Rigotti & Malgieri, 2023). At the same time, user behavior will be constrained by the readiness of the technology, the limitations of the virtual environment, the users involved in the platform, and of course the inevitable implications of the usage of such complex environments. As more people engage in the metaverse, questions around the representation of diverse identities and the potential for discrimination within virtual environments become increasingly relevant.

Hackl et al. (2022) explain that the creation of the avatar in the metaverse will soon become extremely valuable to our sense of self and social acceptance. When users embody an avatar, they feel a sense of ownership and agency over the avatar's body and perceive self-location within its boundaries (Kilteni et al., 2012; Lenggenhager et al., 2007; Slater et al., 2009). This phenomenon is called mediated embodiment and refers to the experience of perceiving an avatar's body as one's own through technology (Aymerich-Franch, 2018). Three components define the illusion of mediated embodiment: body ownership, self-location, and agency (Kilteni et al., 2012; Longo et al., 2008). Body ownership pertains to the feeling of possessing the avatar's body, while self-location is the perception of one's position in space, and agency is the subjective sense of controlling the avatar's actions (Aymerich-Franch & Ganesh, 2015; Gallagher, 2000; Tsakiris, 2010; Blanke & Metzinger, 2009). The experience of mediated embodiment is not limited to human-like avatars and has been observed in non-human entities such as animals and robots (Ahn et al., 2016; Aymerich-Franch et al., 2016, 2017a, 2017b).

The creation of avatars in the metaverse is valuable for our sense of self and social acceptance. People can express themselves freely and break free from socio-physical constraints. Transgender users of the metaverse can create avatars that better reflect their true selves and avoid social marginalization. Similarly, sex workers can separate their online lives from offline work and avoid socio-economic obstacles. However, there might be social pressure for meta-users to "normalize" themselves and fit in with other participants (Burrows, 2022). It could also lead to people conforming to socially accepted norms and experiencing "cosmetic vulnerability" due to aesthetic consumerism (García-Sánchez, 2016). The physical appearance of the human body links closely to societal pressure to conform to accepted norms. Consequently, individuals are increasingly attempting to manage their appearance and abilities to enhance social interaction and gain recognition. Their goal is to present themselves as non-marginalized individuals deserving of social acceptance and inclusion in society. Non-white individuals in predominantly white-designed virtual worlds may lighten their avatar's skin to fit in (something called whitewashing) and avoid discrimination or exclusion; women may choose to present as men to prevent harassment or discrimination within virtual spaces. These choices highlight the way dominant social norms can influence how individuals present themselves within virtual environments. However, it is essential to recognize that these choices are not made in a vacuum and are influenced by larger social forces such as systemic racism, sexism, heteronormativity, homophobia, and transphobia.

The creation of conformist avatars perpetuates the societal pressure to conform to accepted norms of appearance and performance (Rigotti & Malgieri, 2023; Narula, 2022). This pressure has led to physical training, dieting, cosmetic surgery, and the crafting of social media profiles to pursue a desirable appearance. The normativity expected to emerge in the avatar creation process is not limited to protected categories such as gender/sex, race, age, and disability. Other personal characteristics such as breast size, height, and fashion style fetishized in society could also come into play. As a result, individuals may feel the need to create a conformist avatar to address bodily dissatisfaction and social subordination. However, they may also fear how others react to a conformist avatar that does not correspond to their physical being. In the offline world, undergoing cosmetic surgery is often shrouded in secrecy, and deception is common in online and offline dating. In the metaverse, this could result in individuals feeling pressure to create avatars that do not truly reflect their physical appearance or personal characteristics.

All in all, the metaverse's potential for dismantling barriers is notable. However, it raises worries about conforming to appearance norms and avatar-related vulnerabilities, highlighting how offline societal pressures persist online (Rigotti & Malgieri, 2023).

**Existing datasets lack representative data**

Existing datasets have been shown to be limited in terms of providing adequate representation of human diversity. This results from several factors. Data collection may be biased due to inadequate sampling techniques, non-response bias, or lack of representation of certain groups in the population. For example, certain groups, such as low-income households or rural communities, may be underrepresented in surveys or censuses, leading to incomplete data. However, most of the time, the system's designer's sampling technique causes data incompleteness. For instance, in a recent review of databases used for affective computing, some researchers found out that the usual mean age of subjects is almost exclusively between 20 and 30 years of age across datasets, excluding older age groups (50 and up), which are highly underrepresented, probably because many research groups use undergraduate or graduate students from their programs to participate in their studies (Verhoef & Fosch-Villaronga, 2023). Another finding is that general-purpose affective computing datasets do not mention any inclusion of populations of subjects with varying (mental) health conditions. Mental health conditions such as depression or schizophrenia (Gur & Gur, 2022), however, can affect facial expressions and speech, making it difficult to identify and classify emotions accurately.

Careful data collection with subjects in the lab is time-consuming and costly (Hox & Boeije, 2005). It is, therefore, no surprise that more recent datasets are often created by scraping data from the web.

Using the web has apparent advantages, such as the large volume of data that can be collected in this fashion, which will also increase the inclusion of more diverse data sources. However, a major disadvantage is that basic demographic information concerning data subjects is often unavailable, making it hard to measure and correct potential biases (Zimmer, 2010). As a result, data collection efforts may unknowingly be focused on specific areas or groups, leading to an incomplete dataset.

An increasing number of datasets used in AI contain vast amounts of data collected from the web, such as written reviews on Amazon (Blitzer, Dredze & Pereira, 2007; Dredze, Crammer & Pereira, 2008) or IMDB (Maas et al., 2011) for textual sentiment analysis, or images and movies for the bodily gesture and facial expression recognition through Google image search or YouTube (Mollahosseini, Hasani & Mahoor 2017). Since such datasets do not usually involve recruiting test subjects in a lab, there is no demographic information about the people, making it virtually impossible to assess diversity dimensions for these sources.

In addition to addressing problems with data collection, it is equally important to test systems on a diverse set of participants to ensure accuracy in generalizing and equal treatment across demographic differences. Especially with datasets that make extensive use of data downloaded from the internet, as it is essential to identify and mitigate probable biases by testing the resultant technology on a diverse set of users. Given that AI is often used in sensitive and protected domains, including the (mental) healthcare industry, it is reasonable to apply similar standards surrounding diversity and inclusion across clinical trials (Verhoef & Fosch-Villaronga, 2023). In general, the underrepresentation of marginalized communities in the datasets for AI training is often the product of structural, political reasons (Costanza-Chock, 2020). Additionally, the link between AI testing and marginalized communities can also be the product of political decisions, like the alleged choice of the Chinese government to test software on the Islamic minority of Uighurs to profile them better (Mozur, 2019).

Another potential cause of a lack of diversity in the datasets used for AI training results from concerns about data privacy. Many people are hesitant to share their personal data due to worries about privacy and the security of their data, especially for data that might trigger human vulnerabilities in specific contexts (health conditions, racial data, sexual life data, sexual orientation data, etc.). These privacy concerns have also been translated into legal provisions. The legal protection of personal data (especially in the EU) generally prohibits the processing of special categories of personal data, with few circumscribed exceptions accompanied by additional safeguards (Quinn & Malgieri, 2021).

The reluctance to reveal sensitive data coupled with the legal constraints for the processing of these data can lead to incomplete datasets, where already marginalized communities are underrepresented, especially when it comes to sensitive personal information regarding health or ethnicity. As a result, there is a trade-off between privacy and the need for diverse data. Striking a balance between privacy and diversity, then, is essential to address bias and ensure fairness in AI systems (Žliobaitė & Custers, 2016). The proposed EU AI Act is trying to address this tension. Article 10(5) states that "to the extent that it is strictly necessary for the purposes of ensuring bias monitoring, detection and correction in relation to the high-risk AI systems, the providers of such systems may process special categories of personal data" as defined in the GDPR "subject to appropriate safeguards for the fundamental rights and freedoms of natural persons, including technical limitations on the re-use and use of state-of-the-art security and privacy-preserving measures". Considering that these initial provisions might not be sufficient to resolve the tensions between non-discrimination and privacy in the EU law, the discussion will probably continue to evolve in the next few years (Van Bekkum and Zuiderveen Borgesius, 2022). Addressing these issues will require efforts to improve data collection techniques, address privacy concerns, and allocate sufficient resources to data collection efforts.

**Fixed categories in anti-discrimination law do not account for intersectionality**

Anti-discrimination law is generally based on protected categories (race, sex, sexual orientation, political opinion, religion, belief, disability or chronic illness, civil status, age, nationality) and in specific fields (workplace, access to products and services, welfare state rights, and more). This approach is disputable for practical and theoretical reasons. Looking at the practical reasons, building non-discrimination on fixed groups and fields is elusive, considering that new groups and forms of discrimination emerge daily through AI (Wachter, 2020). Often, people are unaware of being victims of disparate treatment or belonging to a targeted group (Taylor, Floridi, & van der Sloot, 2017). In addition, it has been noted that, in AI design, there is often an exclusive focus on some forms of discriminatory biases (e.g., ethnicity or gender) rather than others (Peters, 2022).

The approach of anti-discrimination law based on closed groups and fields also faces the challenge of intersectionality. Individuals can experience multiple forms of discrimination that intersect and compound each other, resulting in unique experiences of oppression (Creshaw, 1990; Nobel, 2018). For example, a disabled transgender person may face discrimination not only based on their disability but also their gender, creating what has been called double jeopardy (Williams, 2014). The intersection of these two identities, for example, may also create a distinct form of discrimination due to the particular social and cultural connotations surrounding the categories, which are not adequately addressed by traditional anti-discrimination laws (Buchanan et al., 2009; Shaw, Chan, and McMahon, 2012).

From the AI design perspective, preventing AI biases through an intersectional approach is problematic. Computing "intersections" of discrimination sources is indeed quite complex. Recently, computer scientists have tried to address this issue by simplifying intersectional discrimination through a "multi-dimensional" approach (Roy, Horstmann & Ntoutsi, 2023). According to the intersectional approaches, the intersection of more sources of discrimination (gender, race, disability) gives rise to a new form of discrimination, which is unique and not based on the mere arithmetical addition of single sources of discrimination. This intersectional discrimination uniqueness is difficult to be automated through a de-biasing process. Generally, and especially in AI fairness, it is not easy to explain and operationalize how "being black and transgender" differs from "being black + being transgender" in each context. That is why, despite some efforts (Wang et al., 2022), the typical solution for AI designers is "multi-dimensionality," where all factors of discrimination are computed in addition to the others (Roy, Horstmann & Ntoutsi, 2023).

To address these challenges, some scholars have proposed alternative approaches. For instance, instead of focusing on fixed categories, a more dynamic method recognizing the fluidity of identity and the intersectionality of discrimination could help address related impacts (Collins, 2015). Others have suggested targeting systemic discrimination and addressing the root causes of inequality (Sampson & Wilson, 2020). However, that may involve more significant structural changes in a particular society and entail shaking deep-rooted power structures (Costanza-Chock, 2020).

Focusing on theoretical reasons, non-discrimination law is often considered to be based on a neo-liberal conception of "equality," where individuals should be treated equally in similar conditions. Scholars have questioned this approach primarily in the last decades. Fineman (2017) criticizes the traditional approach to equality, affirming that what should be considered is individuals' inherent vulnerability and dependence and the inevitable inequality of humans. This reasoning challenges the traditional liberal focus on individual choice. Instead, it emphasizes all individuals' inherent vulnerability and interdependence, recognizing that everyone is vulnerable to harm and discrimination, but some individuals and communities are more vulnerable than others due to systemic factors such as poverty, racism, and homophobia (Arnold, Rebchook, & Kegeles, 2014). This inevitable inequality should be addressed not through traditional categories and formal or substantive ideas of equality but through an approach to social justice that would challenge the liberal reliance on individual choice. It is in this context that the need to "queer" the ethics of AI arises as one way forward to challenge and rethink the normative suppositions and values underlying AI systems.

## Working towards more inclusive AI practices, methods, and approaches

As we delve into the queering of AI, it becomes evident that inclusivity, diversity, and equality principles should extend to the realm of machines, algorithms, and robots (Poulsen, Fosch-Villaronga, & Søraa, 2020). Embracing a queer perspective in AI can help us challenge normative assumptions, dismantle biases, and foster more inclusive AI applications that respect every individual's fundamental rights. Understanding the potential adverse consequences of AI on specific communities requires developers, researchers, and engineers to employ a combination of different methodologies, including technical and socio legal (Zaga et al., 2023). These methodologies aim at leveraging several forms of expertise and tools to assess and mitigate risks proactively and as a joint effort.

**Queering Artificial Intelligence**

Solving the binarism problem in AI (different from discussing this issue) requires rethinking how we categorize and classify people and data and developing more diverse and representative training datasets for AI systems. Implementing inclusive sampling strategies, standardized documentation of demographic factors, and adopting diversity and inclusion guidelines like those used in clinical trials within the Artificial Intelligence (AI) field is imperative to realize the ideals of diversity, equity, and inclusion (Verhoef & Fosch-Villaronga, 2023).

However, actively incorporating data from marginalized and underrepresented communities without a careful thought process may not be sufficient. Understanding the potential adverse consequences of AI on specific communities may require the application of anticipatory methodologies that can help foresee the inadvertent consequences that AI models may have for individual users and their communities. These methodologies may enable stakeholders to identify and address potential harms *before* they occur. An intersectional analysis, for instance, is a critical framework for understanding the complex interplay of multiple social identities and systems of oppression. Applying this methodology to AI involves examining how AI systems may disproportionately impact marginalized communities on intersecting identities such as race, gender, sexuality, disability, or socioeconomic status.

From the design perspective, adopting a queer perspective in AI necessitates approaches centering on intersectionality. This involves understanding and accommodating users' diverse needs, preferences, and identities first, via user engagement or familiarity with interdisciplinary literature (Ovalle et al., 2023), and then using this information in subsequent practices like the diversification of training datasets. We call this *vulnerability-sensitive design*. The discussion on the value-sensitive design of digital technologies has increasingly included participatory governance and multi-stakeholder participation in critical decision-making of technological business models (Costanza-Chock, 2020). In sum, designers and engineers should actively involve representatives from marginalized communities in the design process to ensure their perspectives are considered. This means that every design process should wonder about potential impacts on individuals and groups of individuals. After pre-assessing the impacts (in terms of risks and severity of interference on fundamental rights), it is possible to identify the most impacted categories and involve their representatives in the critical decisions about the design of those technologies (Gilman, 2022).

Additionally, the field of embodied technologies offers new avenues for queering the field. Since most users experience embodied avatars as representations of themselves, they tend to design their avatars to reflect their physical selves (Freeman & Maloney, 2021). However, people may also want their avatars to reveal a different self in online games other than that presented by their physical appearance (Kafai et al., 2010). In any case, research continuously shows the lack of representation of several communities, such as the disabled population, in avatar design on mainstream social VR platforms (Zhang et al., 2022) or in affective computing (Verhoef & Fosch-Villaronga, 2023). These findings reveal that although technology may offer certain affordances with respect to diversity, there is still a long go to making it a reality.

Robots have the potential to challenge normative assumptions about gender, sexuality, and identity and embody diverse bodies and selves (Søraa, 2017). Matsuko Deluxe, a well-known celebrity and advocate for those who are different, such as plus-sized individuals, transgender individuals, and members of the LGBTQIA+ community, serves as a paragon for celebrating uniqueness and fostering inclusivity. Together with Japanese researchers, they created Matsukoroid, a robotic likeness of Matsuko Deluxe that helps reflect on the diversity of human bodies, experiences, and identities. The physical appearance of Matsukoroid and other gendered robots can be intentionally designed to break away from traditional gender stereotypes and offer a more expansive representation of bodies and identities. For example, the robot could have a body shape that is not confined to typical gender norms, with a range of physical attributes that reflect the diversity of human forms (Nomura, 2017). This design flexibility allows individuals, regardless of their gender identity or body type, to relate to and feel represented by the robotic figure. Robots like Matsukoroid can also promote inclusivity through their behavior, interactions, and programming. They can be programmed to use gender-neutral language, respect personal pronouns, and engage in conversations that embrace diverse perspectives, which ensure that the robot's behavior aligns with principles of inclusivity and sensitivity towards different identities.

**Queering the ethics of AI: Intersectionality**

The queering of the ethics of AI is essential to promoting a more inclusive and equitable technological landscape. By adopting a queer perspective in AI ethics, we strive to create a world that embraces diversity in all its forms. This may involve rethinking how we categorize and classify people and data and developing more diverse and representative training datasets for AI systems. However, simply incorporating data from marginalized and underrepresented communities is not enough. Understanding the adverse consequences of AI on specific communities requires the application of anticipatory methodologies that can foresee the inadvertent consequences of these technologies.

An intersectional analysis can help understand the complex interplay of multiple social identities and systems of oppression (Christensen, & Jensen, 2012). Applying this methodology to AI involves examining how AI systems may disproportionately impact marginalized communities based on intersecting identities such as race, gender, sexuality, disability, or socioeconomic status (Ciston, 2019). By considering these communities' unique experiences and vulnerabilities, researchers can identify potential biases, discriminatory outcomes, or exclusionary practices in AI systems proactively.

Participatory design and co-creation methodologies also necessitate active community members, particularly those most likely to be affected, in the design, development, and evaluation of AI systems. This inclusive approach ensures that community values, needs, and concerns are integrated into the decision-making processes, helping to mitigate the risks and negative impacts on marginalized groups.

Ethical impact assessments are systematic frameworks that evaluate emerging technologies' potential implications and consequences. This methodology involves thorough assessments of AI systems to identify potential harms and unintended consequences on communities (Stahl et al., 2022). By examining issues such as privacy, fairness, accountability, and power dynamics, ethical impact assessments can provide insights into the potentially adverse impacts of AI and inform the development of mitigating strategies and policies. Impact assessments should be comprehensive and include an analysis of the societal and ethical implications of new technologies and any potential impact they may have on fundamental rights (whether individual or collective) (Mantelero, 2022).

Unfortunately, traditional non-discrimination discussions tend to focus only on a few fixed categories of discrimination, with restricted ex-post solutions and in limited contexts. The more diverse and complex our society is, the more we should aim at embracing a layered approach to human vulnerabilities (Luna, 2009 and 2019). Users of digital technologies might be not only at the intersection of two or more traditional groups but also unaware of being part of some groups who are particularly vulnerable in specific contexts (Malgieri, 2023). A fluid, open, and layered approach to assessing new technology's risks and harms is necessary, even though every practical solution might be highly contextual and difficult to generalize in terms of a specific one-fits-all ethical framework of reaction.

By employing these anticipatory methodologies, including life-cycle assessments, we can gain a comprehensive understanding of the adverse consequences AI may have on specific communities (Rose et al., 2021; Chiang et al., 2021; Wender et al., 2014). This knowledge allows for proactive measures to address biases, inequalities, and discriminatory outcomes, ensuring that AI technologies are developed and deployed in a way that respect and safeguard all individuals' rights and well-being.

**Queering (non-discrimination, data protection and consumer protection) law**

Our proposal to "queer" the ethics of AI also entails queering the different legal frameworks that affect AI, particularly non-discrimination, data protection, or consumer law. Although our call for queering this field sounds like a mere provocation, we aim to seriously open a debate towards a more fluid and layered approach to regulating AI systems and their effects on individuals and society. There are numerous ways to change the paradigm and overcome many of the issues we have identified (binarism in law and computer science, static approach to discrimination, fallacious neoliberal paradigm in addressing inequalities, legal inefficiencies in governing social oppressions, and privilege-based conformism on social media).

More radical proposals call for a "personalization" of law (Ben Shaar, 2021) to reconsider the concept of equality. Even if we are not advocating for a personalized law, we call for a fluidity of law, where layered subjective perspectives (especially of marginalized and underrepresented individuals and groups) are increasingly considered through a change of paradigm. To combat power imbalances in the AI ecosystem, we should perhaps look for alternative solutions, like ex-ante participation of different stakeholders in the design of AI systems and open-minded approaches to impact assessments of AI, e.g., based on vulnerability theories rather than the traditional legal paradigms of non-discrimination law, consumer protection law, etc.

Algorithm auditing could also help check whether the training data, among other aspects, are inclusive and diverse. Though the financial, aviation, chemical, food, and pharmaceutical industries have effectively included audits to control risk, there has been little to doing the same for AI (Calleja, Drukarch, & Fosch-Villaronga, 2022). Algorithmic auditing involves systematically analyzing and evaluating AI systems to uncover biases, discriminatory patterns, and other adverse consequences (Raji et al., 2020). Auditing techniques can range from statistical analysis to interpretability methods, helping to shed light on how AI systems might reinforce or exacerbate existing social practices and inequalities (Bartley et al., 2021). Using these techniques, it is possible to identify potential harms and inequities that may disproportionately impact specific communities by examining the underlying algorithms, training data, and decision-making processes.

However, it is not enough to identify problematic patterns or biases; auditing processes must lead to tangible and meaningful actions that drive positive change. Actionable algorithmic auditing involves going beyond analysis and actively implementing measures to mitigate harms, promote fairness, and enhance transparency. It requires developing and implementing clear regulations that hold developers of algorithmic systems accountable and creating mechanisms for ongoing monitoring and evaluation. Algorithm auditing, thus, necessitates collaboration between auditors, developers, policymakers, and affected communities to ensure that recommendations are practical, feasible, and inclusive and effectively prevent adverse outcomes ranging from discrimination, privacy invasion, or safety (Raji & Buolamwini, 2019). By emphasizing actionable outcomes, algorithmic auditing can pave the way for the responsible, equitable, and just use of automated decision-making systems.

Conclusion

Technology users are vital in shaping social constructions, relationships, and practices (Douglas, 2012). They engage in activities such as consuming, modifying, domesticating, designing, reconfiguring, and resisting technological advancements. Consequently, when technology is designed solely based on traditional heteronormative perspectives, it risks excluding various minority groups, as it prioritizes the needs and experiences of the majority (Poulsen et al., 2020). Failure to incorporate the intersectional realities, perspectives, and identities of users in the development of AI will result in the persistence of implicit biases. Furthermore, this exclusion will keep many individuals largely invisible, voiceless, marginalized, and unaware of the potential impacts of these technologies on their lives (Criado-Perez, 2019).

As AI will increasingly play a more significant role in shaping complex virtual spaces, it is essential to consider how these technologies can promote more inclusive and equitable representations of diverse identities. This promotion can include using AI to develop more diverse and representative avatars and implementing anti-discrimination policies and practices within virtual environments. Additionally, it is paramount to acknowledge and challenge dominant social norms and power structures that perpetuate discrimination and exclusion in virtual spaces.

The deployment of AI in society will increasingly be an influential factor in shaping individuals' sense of self in the modern world. However, it is crucial to recognize that efforts toward diversity and inclusion in the AI realm are necessary to avoid perpetuating normative views that deny the existence and experiences of everyone, but in particular to specific collectives that have been traditionally marginalized and excluded at best by society, such as transgender community. Addressing this issue requires holistic inclusion strategies that extend to multiple levels, including how these communities can benefit from or can be impacted by AI technology (Poulsen, Fosch-Villaronga & Søraa, 2020). To achieve this, we propose to queer the ethics of AI so that more efforts take place to understand how different communities, including women, LGBTQ+, and persons with disabilities, interact with and value AI technologies. This knowledge can inform the design, creation, and implementation processes, ensuring the meaningful inclusion of these communities.

Queering the ethics of AI goes beyond technical considerations. It requires fostering diversity and inclusivity within the AI research and development community. Encouraging and supporting individuals from underrepresented backgrounds, including queer individuals, to pursue careers in AI can bring unique perspectives and insights to the field. This diversity of voices is crucial in shaping AI systems that cater to a wide range of human experiences and contribute to a more equitable society. Indeed, since AI is not neutral or objective but reflects and amplifies its creators' and users' biases and values, addressing these issues can contribute to creating AI systems that are truly fair, just, and equitable for all.

## Acknowledgement

This paper is part of the Safe and Sound project, a project that has received funding from the European Union's Horizon-ERC program Grant Agreement No. 101076929. Views and opinions expressed are however those of the author(s) only and do not necessarily reflect those of the European Union or the European Research Council. Neither the European Union nor the granting authority can be held responsible for them.